\documentclass[12pt]{article}
\large

\title{ On the local limit of quantum field theories defined on the loop
space.}

\vspace{0.5cm}

\author{ V.V. Belokurov and E.T. Shavgulidze    \\
{\em Lomonosov Moscow State University, Russia }
\\ {\it e-mail: belokur@rector.msu.ru}}

\date{ \ \ \  }
\begin{document}
\maketitle

The local limit of a quantum field theory on the loop space is studied. It is proved that the invariance of the theory with respect to the group of diffeomorphisms leads to Feynman diagrams convergence in the local limit.

\vspace{0.5cm}

In the paper \cite{(Solovyov)} we proposed a quantum field theory model defined on the loop space \cite{(Pressley)}.
Let us remind the idea of the model construction. The momentum space of the theory is the space
$$
\mathcal{P}= C(S^1,\mathbf{R}^4)
$$
of all continuous maps of a circle of a unit length into
$\mathbf{R}^4$ with the norm
$$ \|p\| _{\mathcal{P}} =\|p\| =\max \limits _{ \tau \in S^1}\|p(\tau)\|\,,$$
where $\|\cdot \|$ is the  norm in $\mathbf{R}^4$.

An arbitrary element of this space can be represented in the form
\begin{equation}
   \label{1}
p(\tau)=r+\frac{1}{\sqrt{\lambda}}\xi(\tau)\,.
\end{equation}
Here $\xi(\tau)$ satisfies the condition
\begin{equation}
   \label{2}
\int \limits _{S^{1}}\, \,\xi(\tau)\, d\tau\,=\,0\,.
\end{equation}
In the "local" limit $(\lambda\rightarrow +\infty )\,,$
$ p(\tau)$ turns into $\,r\,$
( the point in $\mathbf{R}^4$ ).

The group of
diffeomorphisms of a circle
$$
 G=Diff\,^{2}_{+}\left( S^{1}\right)\,,
\ \ 
g\in G \ \ \ \{g:\ S^{1}\longrightarrow S^{1}\,,\ \ \
g'(\tau)>0\}
$$
acts on the space $\mathcal{P}$ in the
following way:
\begin{equation}
   \label{3}
gp(\tau)=p\,\left(g^{-1}(\tau)\right)
\,\frac{1}{\sqrt{\left(g^{-1}\right)'(\tau)}}\,.
\end{equation}
On the space $\mathcal{P}$ there are no measures invariant with respect to the group $G$.
However, the Wiener measure
\begin{equation}
   \label{4}
w_{\lambda}(dp)=
\exp\left\{-\frac{\lambda}{2}\,\int \limits
_{S^{1}}\, \,\|p'(\tau)\|^{2}\, d\tau \right\}\,dp
\end{equation}
is
quasi-invariant \cite{(Shavgulidze1978)} and  transforms as
\begin{equation}
   \label{5}
  w_{\lambda}(d\,(gp)\,)=
  \exp\left\{\frac{\lambda}{4}\,\int \limits
_{S^{1}}\,\mathcal{S}_{g}(\tau) \,\|p(\tau)\|^{2}\, d\tau
\right\}\ w_{\lambda}(dp)\,.
\end{equation}
 Here $\mathcal{S}_{g}$ denotes the
Schwarz derivative
\begin{equation}
   \label{6}
\mathcal{S}_{g}(\tau)=
\left(\frac{g''(\tau)}{g'(\tau)}\right)'
-\frac{1}{2}\left(\frac{g''(\tau)}{g'(\tau)}\right)^2\,.
\end{equation}
Note that
$
w_{\lambda}(dp)=dr\ w_{1}(d\xi)\ .
$

Let $E$ be the Hilbert space  of all square-integrable over the Wiener
measure functions $\varphi :\mathcal{P}\to \mathbf{C}$ with the conjugation rule
$$
\overline{\varphi (p)}=\varphi(-p)\,.
$$

Functions $\varphi$ realize a regular unitary
representation of the group $G$ in the Hilbert space $E$:
\begin{equation}
   \label{7}
   g\,\varphi (p)\,=
   \,\varphi (g^{-1}\,p)\
\exp\left\{\frac{\lambda}{8}\,\int \limits
_{S^{1}}\,\mathcal{S}_{g^{-1}}(\tau) \,\|p(\tau)\|^{2}\, d\tau
\right\}\,.
\end{equation}
The scalar product
$
\int \limits_{\mathcal{P}}\ \varphi(p)\overline{\phi(p)\,}
\,w_{\lambda}(dp)
$
being invariant.

The free action is of the form
 \begin{equation}
   \label{8}
\mathcal{A}_0^{g}[\varphi ] =\int \limits_{\mathcal{P}}
\, \left (\int\limits_{S^1}\|g\,p(\tau ) \|^2\, d
\tau + m^2\right) \,|\varphi (p) |^2\ w_{\lambda}(dp)\,
\end{equation}
 with $g$
being fixed.

For continuous field functions $\varphi (p) $ in the "local" limit we get
\begin{equation}
   \label{9}
\lim \limits_{\lambda\rightarrow +\infty}\,\mathcal{A}_{0}^{g}[\varphi
]=
\int\limits_{S^1}(g '(\tau )) ^2\, d
\tau\,\int \limits_{\mathbf{R}^4} \,|\varphi (r) |^2\,
\left(\|r \|^2 + m^{2} \right)\, dr\,.
\end{equation}
That is the action of free scalar field multiplied by the factor that depends on $g\,.$

Let us also consider  the following interaction term.
$$
\mathcal{A}_1^{g}[\varphi ] =\int\limits_{\mathcal{P}}\cdots
\int\limits_{\mathcal{P}}\, \left (\,\int\limits_{S^1} \delta (g\, p_1(\tau)+ g\,p_2(\tau)+ g\,p_3(\tau)+ g\,p_4(\tau))\, d \tau \,\right)\,
$$
\begin{equation}
   \label{10}
\varphi (p_1)\varphi (p_2)\varphi
(p_3)\varphi (p_4)\ w_{\lambda}(dp_1)\,w_{\lambda}(dp_2)\,w_{\lambda}(dp_3)\,w_{\lambda}(dp_4)\,.
\end{equation}

For continuous field functions $\varphi (p) $ in the "local" limit it gives the usual interaction
$\varphi^{4}$ multiplied by the factor that depends on $g\,.$
$$
\lim \limits_{\lambda\rightarrow +\infty}\,\mathcal{A}_1^{g}[\varphi ] =
\int\limits_{S^{1}}\,\frac{d\tau}{g'(\tau)}\int\limits_{\mathbf{R}^4}\cdots
\int\limits_{\mathbf{R}^4}\,\delta ( r_1+ r_2+ r_3+ r_4)\,
$$
\begin{equation}
   \label{11}
 \varphi (r_1)\varphi (r_2)\varphi
(r_3)\varphi (r_4)\ dr_1\, dr_2\, dr_3\, dr_4\, .
\end{equation}

It is convenient to make the change of variables
$$
q(\tau)=g\, p(\tau)\,, \ \ \psi(q)=g\,\varphi(q)\,.
$$
The new momentum variable $q$ has the form
$\ q(\tau)=\rho+\frac{1}{\sqrt{\lambda}}\eta(\tau)\,.$

In the new variables the free action 
\begin{equation}
   \label{12}
\mathcal{A}_0^{g}[\psi ] =\mathcal{A}_0[\psi ] =\int \limits_{\mathcal{P}}
\, \left (\int\limits_{S^1}\|q(\tau ) \|^{2} d
\tau + m^2 \right ) \,|\psi (q) |^2\,w_{\lambda}(dq)\,,
\end{equation}
does not depend on $g$.
      
      The dependence of $\mathcal{A}_1^{g}[\psi ]$  on $g$ is only in the Schwarz derivative:
$$
\mathcal{A}_1^{g}[\psi ] =\int\limits_{\mathcal{P}}\cdots
\int\limits_{\mathcal{P}}\, 
\left (\,\int\limits_{S^1} \delta ( q_1(\tau)+\ldots+ q_4(\tau))\, d \tau \,\right)\,
$$
$$
\exp\left\{\frac{\lambda}{8}\,\int \limits_{S^{1}}\,\mathcal{S}_{g^{-1}}(\tau) \, \left ( \, \|q_{1} (\tau)\|^{2}+\ldots+\|q_{4}(\tau)\|^{2} \, \right ) \, d\tau
\right\}\,.
$$
\begin{equation}
   \label{13}
\psi (q_1)\psi (q_2)\psi
(q_3)\psi (q_4)\ w_{\lambda}(dq_1)\,w_{\lambda}(dq_2)\,w_{\lambda}(dq_3)\,w_{\lambda}(dq_4)\,.
\end{equation}

To get the coincidence with the action of the ordinary theory  when $
\lambda\rightarrow +\infty $ we will consider only the diffeomorphisms $g_{\lambda}(\tau)$ that satisfy the conditions
\begin{equation}
   \label{14}
\lim \limits_{\lambda\rightarrow +\infty}\ g''_{\lambda}(\tau)=0\,,\ \lim \limits_{\lambda\rightarrow +\infty}\ g'_{\lambda}(\tau)=1\,.
\end{equation}
Such diffeomorphisms turn into identity in the limit case
$$\lim \limits_{\lambda\rightarrow +\infty}\ g_{\lambda}(\tau)= \tau\,.$$

Now we can substitute
\begin{equation}
   \label{15}
\frac{g''(\tau)}{g'(\tau)}=\frac{1}{\sqrt{\lambda}}\,f(\tau)\,.
\end{equation}
Here,
$
\ f\in C(S^1, \mathbf{R}):\,
\int\limits_{S^1}f(\tau)\,d\tau=0\,.
$

In terms of $f$ the Schwarz derivative $\mathcal{S}_{g}$ takes the form
\begin{equation}
   \label{16}
\mathcal{S}^{f}(\tau)=\frac{1}{\sqrt{\lambda}}\,f'(\tau)-\frac{1}{2\lambda}\,f^{2}(\tau)\,.
\end{equation}

We average the interaction term over $f$ using the quasi-invariant Wiener measure $w_{\alpha}(df)$
\cite{(Shavgulidze2000)}
\begin{equation}
   \label{17}
\mathcal{A}_1\,=\,\int \mathcal{A}_1^{g}\,w_{\alpha}(df)\,.
\end{equation}

Note that the integrand in the averaged action $\mathcal{A}_1\,$ for every $g$
has the same "local" limit
(ordinary interaction $\varphi^{4}$).

Thus, in the "local" limit the equations (\ref{12}) and (\ref{17}) lead to the standard classical action for the scalar field. 

However in quantum theory the situation is quite different. 
In quantum theory we integrate over the space of field functions $\psi \,.$ But even the Gaussian measure
$$\exp \{-\mathcal{A}_0 [\psi]\}\,d\psi$$
 is concentrated in the space of discontinuous functions. That is why to take the naive "local" limit
$$
\psi(q)\rightarrow \psi (\rho )
$$
 in the integrand is not correct.

The correct way is to integrate over $\psi$ first and then to pass to the limit $\lambda\rightarrow +\infty\,.$
In the limit, loops contract to points and diffeomorphisms turn to identity.

 But in the subsequent calculations the memory about the loop space remains and yields Feynman diagrams convergence.

We illustrate it in the case of the simplest diagram (so called "fish").
The functional integral for this diagram is of the form
\begin{equation}
   \label{18}
\int \psi (q_{1})\,\psi (q_{2})\,\psi (q_{3})\,\psi(q_{4})\,\left(\mathcal{A}_1 [\psi]\right)^{2}\exp\{-\mathcal{A}_0[\psi ]\}\ d\psi\,.
\end{equation}
Performing the integration with the help of the equation
$$
\int \int \psi (q_{1})\,a(q_{1})\,w_{\lambda}(dq_{1})\,\int\psi (q_{2})\,b(q_{2})\,w_{\lambda}(dq_{2})
\exp\{-\mathcal{A}_0[\psi ]\}\ d\psi\,=
$$
\begin{equation}
   \label{19}
\int \frac{1}{\Omega (q)}\,a(q)b(-q)\,w_{\lambda}(dq)\,,\ \ \ (\,\Omega(q)=\|q\|^{2}+m^{2}\,)
\end{equation}
we get
$$
\int\  \left[\int
\delta (q_{1}(\tau)+ q_{2}(\tau)- q_{5}(\tau)- q_{6}(\tau))d\tau \right]
\left[ \int \delta (q_{3}(\tau)+ q_{4}(\tau)+ q_{5}(\tau)+ q_{6}(\tau))d\tau \right]
$$

$$
\exp\left\{\frac{\lambda}{8}\,\int \mathcal{S}^{f_{1}}(\tau) \, \left (  \|q_{1} (\tau)\|^{2}+\|q_{2} (\tau)\|^{2}+\|q_{5} (\tau)\|^{2}+\|q_{6} (\tau)\|^{2}  \right ) \, d\tau
\right\}
$$
$$
\exp\left\{\frac{\lambda}{8}\,\int \mathcal{S}^{f_{2}}(\tau) \, \left (  \|q_{3} (\tau)\|^{2}+\|q_{4} (\tau)\|^{2}+\|q_{5} (\tau)\|^{2}+\|q_{6} (\tau)\|^{2}  \right ) \, d\tau
\right\}
$$
\begin{equation}
   \label{20}
\frac{1}{\Omega (q_{5})}\frac{1}{\Omega (q_{6})}w_{\lambda}(dq_{5})\,w_{\lambda}(dq_{6})\,w_{\alpha}(df_{1})\,w_{\alpha} (df_{2})\,.
\end{equation}

Now the limit $\lambda\rightarrow
+\infty $ results in
$$
\delta (\rho _{1}+\cdots +\rho _{4})
$$
$$
\int \exp \left\{ -\frac{1}{16}\left(\|\rho\|^{2}+\| \rho _{1}+ \rho_{2}- \rho \|^{2} \right)
\int \,(f_{1}^{2}(\tau)+f_{2}^{2}(\tau))\, d\tau\, \right\}
$$
$$
\exp \left\{\frac{1}{4} \int \left(f'_{1}(\tau)+f'_{2}(\tau )\right)  \left[(\rho,\eta_{5}(\tau )+(\rho_{1}+ \rho_{2}- \rho,\eta_{6})
\right]\, d\tau \right\}
$$
\begin{equation}
   \label{21}
J \,\frac{1}{\Omega (\rho )}\frac{1}{\Omega (\rho _{1}+ \rho _{2}- \rho )}\,d\rho\,w_{1}(d\eta _{5})\,w_{1}(d\eta _{6})\,w_{\alpha}(df_{1} )\,w_{\alpha}(df_{2} )\,.
\end{equation}
Here, the factor $J$ does not depend on $ \rho\,, \eta _{5}\,,\eta _{6}\,. $

Integrations over $\eta_{5},\,\eta_{6}$ and $f_{1}+f_{2}$
result in
\begin{equation}
   \label{22}
I=
\int \frac{1}{\Omega (\rho)} \frac{1}{\Omega (\rho_{1}+ \rho_{2}- \rho)}\,\exp \left\{ -\frac{1}{16}\|\rho\|^{2}
\int\,(v^{2}(\tau)\,d\tau\, \right\}\,d\rho\,w_{\alpha}(dv)\,,
\end{equation}
where $v=\frac{1}{\sqrt{2}}(f_{1}-f_{2})\,.$

From the simple estimations it follows that the integral $I$ is convergent:
$$
I\,\leq C_{1}\,\int\,
\exp \left\{ -\frac{1}{16}\|\rho\|^{2}
\int\,(v^{2}(\tau)\,d\tau\, \right\}\,d\rho\,w_{1}(dv)\,\leq
$$
\begin{equation}
   \label{23}
C_{2}\int\,\frac{1}{\left(\int \,v^{2}(\tau)\,d\tau \right)^{2}}\,\exp\left\{-\frac{\alpha}{2}\int (v'(\tau)^{2}d\tau\right\}\,dv\,<\,\infty \,.
\end{equation}

Due to the averaging over $f$ in the action (\ref{17}) the same "subtraction" mechanism is valid for other diagrams. After integration over $\eta_{i}$ every line gets the decreasing  factor $\exp \left\{ -\frac{1}{32}\|\rho\|^{2}
\int\,(v_{i}^{2}(\tau)\,d\tau\, \right\}$ that ensures the convergence of a diagram.

All that will not look so surprising if we consider the following example that is in some sense analogy with the
proposed model.

The integral over the real half-axis
$$
\lim \limits_{\lambda\rightarrow +\infty}\,\int\limits_{0}^{\lambda}\,\frac{x\,e^{\imath\sigma}}{\left(x\,e^{\imath\sigma}-1\right)^{2}}\,dx
$$
diverges.

However, the integral over infinitely narrow strip in $\sigma $ exists:
$$
\lim \limits_{\lambda\rightarrow +\infty}\,\int\limits_{-\frac{1}{\lambda}}^{\frac{1}{\lambda}}\,d\sigma\,
\int\limits_{0}^{\lambda}\,\frac{x\,e^{\imath\sigma}}{\left(x\,e^{\imath\sigma}-1\right)^{2}}\,dx\,=2\pi\,.
$$

Note that the averaging over the Wiener measure $w_{\alpha}(df)$ contains an arbitrariness in the choice of the dimensional
parameter $\alpha \ (\,[\alpha]=[p^{2}]\,)\,.$

The connection of this arbitrariness with the known arbitrariness in the subtraction point in the standard theory as well as the connection of the proposed model with the ordinary renormalized quantum field theory are now under study.

\end{document}